\documentclass[aps,prd,showpacs,amsmath,nofootinbib,
superscriptaddress,twocolumn,preprintnumbers]{revtex4}
\usepackage{natbib}
\usepackage{amsmath}
\usepackage{amssymb}
\usepackage{graphicx}
\usepackage{color}
\usepackage{amsmath}
\usepackage{amsthm}
\usepackage{slashed}
\usepackage{soul}
\usepackage[usenames,dvipsnames,svgnames,table]{xcolor}
\begin{document}

\title{Conformal Higgs Gravity}

\author{Meir Shimon}
\affiliation{School of Physics and Astronomy, 
Tel Aviv University, Tel Aviv 69978, Israel}
\email{meirs@wise.tau.ac.il}

\title{Conformal Higgs Gravity}

\begin{abstract}
It is shown that gravitation naturally emerges 
from the standard model of particle physics if {\it local} scale invariance 
is imposed in the context of a single conformal (Weyl-symmetric) theory. 
Gravitation is then
conformally-related to the standard model via a conformal transformation, merely a 
function of the number of fermionic particles dominating the energy density associated 
with the ground state of the physical system. Doing so 
resolves major puzzles afflicting the standard models of particle physics and cosmology, 
clearly indicating these to be artifacts stemming from {\it universally} employing 
the system of units selected {\it here and now}. In addition to the three known 
fundamental interactions mediated by gauge bosons, a scalar-tensor interaction is also 
accommodated by the theory; its inertial and gravitational sectors are characterized by 
whether contributions to the Weyl tensor vanish or are finite, respectively.
In this approach both inertia and gravity are viewed 
as collective phenomena, with characteristic gravitational Planck scale 
devoid of fundamental meaning; consequently, mass hierarchy and 
Higgs mass instability concerns are avoided altogether. 
Only standard model particles gravitate; dark matter and dark energy 
have an inertial origin, and since 
the Higgs field does not interact with photons it is an ideal candidate for explaining 
the dark sector of cosmology. On cosmological scales the dynamical vacuum-like Higgs 
self-coupling accounts for dark energy, and its observed proximity at present to the energy 
density of nonrelativistic matter is merely a consistency requirement. Spatially varying 
vacuum expectation value of the Higgs field could likely account for 
the apparent cold dark matter on both galactic and cosmological scales.
\end{abstract}


\maketitle

\section{Introduction}

The standard model (SM) of particle physics is 
clearly a remarkable achievement in its success to describe three 
fundamental interactions in a single theory. 
The relatively simple framework rests on a fundamental 
$SU(3)\times SU(2)\times U(1)$ symmetry group. 
The fourth interaction, gravitation, is thought to be well-understood on 
macroscopic scales down to a few $\mu m$, with its dynamical impact increasingly 
more prominent on astrophysical and cosmological scales.

The huge dynamical range spanned by the standard 
particle and cosmological models, over $O(61)$ orders of magnitude,  
across the Planck, Higgs, and Hubble scales, 
leads to one of the most vexing problems in theoretical physics: 
The stupendously delicate fine-tuning of (what seems to be a) 
vacuum energy cancellation to one part in $10^{122}$ 
on cosmological scales, [1, 2], 
as compared to naive expectations for the energy density 
of the vacuum, a fine-tuning known as the `cosmological constant problem', 
e.g. [3].
Another difficulty with the cosmological constant is the fact that its energy density 
is comparable to that of ordinary matter at the present epoch, 
although their respective cosmic histories are very different; 
this seems to require very special initial conditions.

A few other naturalness problems in cosmology revolve around the same 
central question of why is the universe so large and old, as compared to the 
corresponding Planck scales, which are presumed to be `natural' in gravitation. 
Whereas the standard cosmological model fits current measurements remarkably well, 
the {\it essence} of dark energy (DE) and cold dark matter (CDM) -- two key ingredients 
in the model that determine the background evolution, large scale structure 
formation history, and gravitational potential on galactic scales 
-- remain elusive. In addition, a few long-standing 
anomalies afflict the standard cosmological model on the largest observable 
scales. The first to be recognized is the lack of correlations in the 
cosmic microwave background (CMB) anisotropy on angular scales $\gtrsim 60^{\circ}$, 
e.g. [4].

Current understanding of the very early universe
(energies TeV and higher) lacks direct experimental
confirmation. A major underpinning of standard cosmology
is the primordial inflationary phase thought to have
ended at energies as high as $O(10^{16})$ GeV, e.g. [5-8].
A few fine-tuning (naturalness) and conceptual problems
generically afflict inflation but its exceptional
role in explaining and predicting a variety of cosmological
phenomena is indeed remarkable. 

One notable fine-tuning in the SM is the `hierarchy problem', i.e. the 
exceedingly small Higgs mass compared to the `natural' (Planck 
scale) ultraviolet (UV) cutoff on virtual momenta that could potentially undermine 
the entire spontaneous symmetry breaking (SSB) mechanism. 
The Planck energy scale is believed 
to be the regime in which quantum gravitational effects are dominant, although 
a satisfactory theory of quantum gravity has not yet been formulated. 
Taming the quadratic UV corrections to the Higgs mass is hoped to be achieved 
with, e.g. supersymmetric extensions of the SM. 
In addition, supersymmetry holds the promise of providing the 
theoretical basis for a slew of new particles that could account 
for CDM. However, at present there is no 
experimental evidence for supersymmetry, 
nor DM particles have been ever detected.

The only characteristic energy scale in the SM is the vacuum 
expectation value (VEV) of the Higgs field that endows 
masses to elementary particles. 
In the absence of this scale the SM is {\it globally} 
scale-invariant in Minkowski spacetime. 
Breakdown of the chiral symmetry of quantum chromodynamics (QCD) 
is likewise responsible for hadron masses. 

The main objective of this work is to demonstrate the viability of an alternative 
physical framework based on a Weyl-symmetric scalar-tensor theory, that 
essentially unifies the SM and gravitation under the same framework. 
We argue that {\it Weyl symmetry manifests the invariance of the three 
fundamental interactions to the number of fermionic particles in a given system} 
(and can thus be described in terms of single-particle ground states). This {\it classical} 
symmetry applies to the ground state of a given physical system in particular, 
in the sense that gravitation pertains to a multi-particle ground state of a given system, 
whereas the three fundamental (`inertial') interactions are described by 
one-particle ground states.
We identify this as the root cause for the apparent DE problem in 
cosmology, mass hierarchy in the SM, and other related problems. 
This unification scheme provides important new insight into the 
nature of gravity, the essence of mass, problematic aspects of electroweak 
(EW)--Planck hierarchy, the closely-related Higgs mass instability, DE, and CDM. 
We adopt the units convention $\hbar=1=c$ throughout, and the metric signature is 
mostly negative.

\section{Theoretical Framework}

The SM, minimally-coupled to metric field, is described by the following action
\begin{eqnarray}
\mathcal{I}_{SM}&=&\int\left[\mathcal{L}_{M}(|H|,\psi,A^{\mu},g_{\mu\nu})
+D_{\mu}H^{\dagger} D^{\mu}H\right.\\
&+&\left.\frac{1}{2}m_{H}^{2}H^{\dagger}H-\lambda_{SM}(H^{\dagger}H)^{2}\right]
\sqrt{-g}d^{4}x,\nonumber 
\end{eqnarray}
where a classically-irrelevant constant term $\propto m_{H}^{4}$ 
(which guarantees that the potential energy associated with the Higgs field at the 
ground-state vanishes, consistent with the weak gravitational field on, e.g. the face 
of Earth) has been omitted. 
The Higgs mass parameter, $m_{H}\approx 125 GeV$, is related to the VEV 
of the Higgs field, $v\approx 246 GeV$, by 
$m_{H}=\sqrt{2\lambda_{SM}}v$. 
Here, $H$, $A_{\mu}$, $\psi$ and $g_{\mu\nu}$ are the Higgs, 
gauge, Dirac and metric fields respectively, $\lambda_{SM}$ is 
the dimensionless Higgs self-coupling parameter, 
and $D_{\mu}H\equiv H_{\mu}+igA_{\mu}H$ is the gauge-covariant derivative of $H$. 
The partial derivative of $H$ with respect to $x^{\mu}$ 
is $H_{\mu}\equiv H_{,\mu}$, dimensionless (gauge) 
coupling constants are collectively denoted $g$, 
and $\mathcal{L}_{M}$ is the lagrangian density of the non-pure-Higgs field sector 
(assumed to depend on $H$ only through its modulus, $|H|\equiv\sqrt{H^{\dagger}H}$, 
via Yukawa couplings of 
the form $\lambda_{i}\bar{\psi}H\psi+h.c.$ where $\lambda_{i}$ is the 
dimensionless Yukawa coupling of the i'th species). 
The SSB term, $\frac{1}{2}m_{H}^{2}H^{\dagger}H$, determines the EW 
scale. While the action is gauge- and diffeomorphism-invariant, 
it is not Weyl-symmetric. 

Clearly, as it stands, Eq. (1) does not account for the gravitational 
interaction. In particular, the fields are only minimally-coupled 
to gravity and the metric tensor is consequently 
{\it non-dynamical}. The Einstein field equations that govern 
general relativity (GR) are obtained from variation of the 
Einstein-Hilbert (EH) action 
$\mathcal{I}_{EH}=\int[\frac{1}{2}m_{p}^{2}(R-2\Lambda)
+\mathcal{L}_{M}]\sqrt{-g}d^{4}x$ with 
respect to the metric field, where 
$m_{p}\equiv(8\pi G)^{-1/2}$, $G$, 
$R$, and $\Lambda$ are the reduced Planck mass, the gravitational 
constant, the curvature scalar, and cosmological constant, respectively. 
The coupling parameter $\frac{1}{2}m_{p}^{2}$, relating 
$\mathcal{L}_{M}$ to spacetime curvature $R$, 
has been set by the requirement that GR reduces to Newtonian 
gravity in the weak field limit, in which $G$ was experimentally determined. 

Generalizing Eq. (1) by adding a conformal coupling of the Higgs 
field to $R$ that {\it replaces} 
the $\frac{1}{2}m_{H}^{2}H^{\dagger}H$ term, 
we obtain the {\it effective} `conformal standard model' (CSM)
\begin{eqnarray}
\mathcal{I}_{CSM}&=&\int\left[\frac{1}{6}H^{\dagger}H R+D_{\mu}H^{\dagger} D^{\mu}H
-\lambda_{SM}(H^{\dagger}H)^{2}\right.\nonumber\\
&+&\left.\mathcal{L}_{M}(H,\psi,A^{\mu},g_{\mu\nu})\right]\sqrt{-g}d^{4}x, 
\end{eqnarray}
where here $\mathcal{L}_{M}$ contains the Yukawa coupling term 
(which would correspond to the lagrangian density of NR gas of 
fermions), 
and $\psi$ is the Dirac field of either fundamental particles like the electron or 
composite particles such as the proton. 
The pre-factor $1/6$ in the curvature term [9] guarantees 
the conformal invariance (up to a {\it classically irrelevant} boundary term) 
of Eq. (2) under {\it local} re-scaling, 
$H\rightarrow H\Omega^{-1}$, $\psi\rightarrow \psi\Omega^{-3/2}$, 
$g_{\mu\nu}\rightarrow g_{\mu\nu}\Omega^{2}$, and 
$\mathcal{L}_{M}\rightarrow\mathcal{L}_{M}\Omega^{-4}$ 
(and consequently $\rho_{M}\rightarrow\rho_{M}\Omega^{-4}$), 
while the vector fields 
$A_{\mu}$ are unchanged (and thus $A\equiv\sqrt{A_{\mu}A^{\mu}}\propto\Omega^{-1}$ 
scales by its physical dimension), and $\Omega(x)$ is an arbitrary function of spacetime.
In other words, all {\it fundamental fields} have well-defined conformal 
weights, i.e. the scalar and Dirac fields have weights $-1$ \& $-3/2$, respectively, 
and the contravariant vector and metric fields have the weights $-2$ \& $0$, respectively. 
The spacetime coordinates are dimensionless in this formalism and are therefore 
invariant to the Weyl transformation. In other words, while coordinate distance 
is Weyl-invariant, metric distance, $\sqrt{g_{\mu\nu}x^{\mu}x^{\nu}}$, transforms 
$\propto\Omega$. The Levi-Civita-based spin 
connection $\Gamma_{\mu}$ appearing in the kinetic term associated with the Dirac field, 
i.e. $\bar{\psi}\gamma^{\mu}(x)(\psi_{\mu}+\Gamma_{\mu}\psi+igA_{\mu}\psi)$, 
where $\gamma^{\mu}(x)$ are Dirac matrices on curved spacetimes, 
e.g. [10], renders the kinetic term of the Dirac 
field of a well-defined conformal weight $-4$. The mass term, 
$|H|\bar{\psi}\psi$, is easily seen to be of conformal weight $-4$. 
All this is necessary for $\mathcal{L}_{M}$ to scale $\propto\Omega^{-4}$.
Similarly, whereas the kinetic term appearing in Eq. (2) 
does not transform as a power of $\Omega$, the combination 
$\frac{1}{6}H^{\dagger}H R+D_{\mu}H^{\dagger} D^{\mu}H$ 
scales proportional to $\Omega^{-4}$. The appearance 
of $R$ \& $\Gamma_{\mu}$ (i.e. derivatives of the metric) 
in Eq. (2) implies that the metric field is now dynamical. Thus, by simply imposing 
Weyl-invariance on the SM (specifically, on the kinetic terms associated with the Higgs 
and Dirac fields), while carrying 
out the necessary modifications, gravitation becomes a {\it dynamical} field.
Quantum mechanical aspects of Eq. (2) (albeit with our Higgs field replaced 
with a general scalar field $\phi$ and with no self-coupling term) have been 
explored in, e.g. [11].  

Thus, perhaps counter-intuitively 
at first glance, an {\it effective} mass term (which is actually a curvature term), 
$\frac{1}{6}H^{\dagger}HR$, is 
introduced in the massless SM 
to restore conformal invariance. In this way, not only that the effective mass term 
does not break conformal invariance, but is also essential for its existence. 

{\it The dynamical VEV of the Higgs field is the local (inverse) radius of curvature} 
($R\equiv 12l_{curv}^{-2}$) -- the locally (on Earth) 
measured mass parameter $m_{H}$ that in the SM is 
taken to be universal {\it by convention}.
Here, it is assumed that the dimensionless ratios of the various 
SM particle properties 
are unchanged in going from Eq. (1) to Eq. (2); all dimensionless 
Yukawa coupling parameters are exactly the same as in the SM. 
This property of the model is equivalent to making the 
requirement that, e.g., quark masses 
given in electron mass units have the same values as in the SM. 
Similarly, the QCD vacuum scale, determined by 
$\langle\bar{\psi}\psi\rangle\rightarrow \Omega^{-3}\langle\bar{\psi}\psi\rangle$, 
or is otherwise related to the VEV of $F_{\mu\nu}F^{\mu\nu}$ 
(where $F_{\mu\nu}$ is the field strength associated with QCD) 
varies in spacetime exactly as the Higgs scale does, and their dimensionless 
ratio is a fixed constant, implying that, e.g., the electron-to-proton mass ratio 
is fixed at its SM value. Since no fixed dimensional quantity is 
allowed in a Weyl-symmetric theory, the CSM cannot contain terms of mass dimension 
different than four, in particular no pseudoscalar fields can couple to 
the field strength in the form 
$\frac{\phi}{f_{a}} F_{\mu\nu}\tilde{F}^{\mu\nu}$ where $\phi$ 
is, e.g., the Peccei-Quinn axion and $f_{a}$ is some fixed energy scale.

Variation of Eq. (2) with respect to $g_{\mu\nu}$ and $H^{\dagger}$ results 
in the following generalized Einstein-like equations and scalar field equation 
\begin{eqnarray}
\frac{HH^{\dagger}G_{\mu}^{\nu}}{3}&=&T_{M,\mu}^{\nu}
-\Theta_{\mu}^{\nu}+\lambda_{SM}\delta_{\mu}^{\nu}(HH^{\dagger})^{2}\\
\frac{H R}{6}-\Box H&+&2\lambda_{SM}(H^{\dagger}H)H
+\frac{\partial\mathcal{L}_{M}}{\partial H^{\dagger}}=0,
\end{eqnarray}
where $G_{\mu}^{\nu}$ is the Einstein tensor, 
and the generalized energy momentum (non-) conservation 
then follows, e.g. [12, 13]
\begin{eqnarray}
T_{M,\mu;\nu}^{\nu}&=&\mathcal{L}_{M,H}H_{\mu}
+\mathcal{L}_{M,H^{\dagger}}H^{\dagger}_{\mu}.
\end{eqnarray}
The effective energy-momentum tensor $\Theta_{\mu}^{\nu}$ 
associated with the Higgs field is
\begin{eqnarray}
\Theta_{\mu}^{\nu}&\equiv&\frac{1}{3}\delta_{\mu}^{\nu}(H^{\dagger}\Box H+H\Box H^{\dagger}
-H^{\dagger}_{\rho}H^{\rho})\nonumber\\
&+&\frac{1}{3}(2H^{\dagger}_{\mu}H^{\nu}+2H_{\mu}H^{\dagger\nu}
-H^{\dagger}H_{\mu}^{\nu}-HH_{\mu}^{\dagger\nu}).
\end{eqnarray}
Here, $f_{\mu}^{\nu}\equiv(f_{,\mu})^{;\nu}$, with $f_{;\mu}$ 
denoting covariant derivatives of $f$, the covariant Laplacian is $\Box f$, 
and $(T_{M})_{\mu\nu}\equiv\frac{2}{\sqrt{-g}}\frac{\delta(\sqrt{-g}\mathcal{L}_{M})}
{\delta g^{\mu\nu}}$ is the energy-momentum tensor.
Eq. (5), which is not independent of (3) \& (4), 
implies that the matter energy-momentum 
is generally not conserved, which is indeed expected 
when $\Lambda$, or particle masses, are spacetime-dependent. 

A straightforward calculation verifies that the 
field equations, Eqs. (3) \& (4), are invariant to Weyl transformations. 
Both $R$ and $G_{\mu}^{\nu}$ do not have well-defined conformal weights, but 
the combinations $\frac{HR}{6}-\Box H$ and $\frac{H^{\dagger}HG_{\mu}^{\nu}}{3}+\Theta_{\mu}^{\nu}$ 
have conformal weights $-3$ \& $-4$ respectively. It is trivial to see that $T_{M,\mu}^{\nu}$ 
and $\frac{\partial\mathcal{L}_{M}}{\partial H^{\dagger}}$ have conformal weights $-4$ \& $-3$, 
respectively. Therefore, any Weyl mapping of classical fields configuration 
[which is consistent with Eqs. (3) \& (4)] is a solution of these field equations.

Since $ds=m\int\sqrt{g_{\mu\nu}dx^{\mu}dx^{\nu}}$, 
the action associated with a massive point-particle of mass $m$, 
is evidently invariant to the 
transformations $g_{\mu\nu}\rightarrow\Omega^{2}g_{\mu\nu}$ and  
$m\rightarrow m/\Omega$. Therefore, the geodesics obtained from its 
variation with respect to coordinates are invariant as well; variation in the mass of 
test particles offset that of the metric so as to leave geodesic motion invariant 
to the mass/length units used. 
Since any GR solution of Eqs. (3) \& (4) can be mapped to a solution 
with dynamical $H$ (but not the other way around) planetary motions are unchanged 
with respect to those obtained in GR. Moreover, because 
the effective metric $g^{eff}_{\mu\nu}\equiv m^{2}g_{\mu\nu}$ 
is invariant, so is its perturbation, i.e. the Newtonian potential $\Phi$, 
and its gradient, i.e. the surface gravity associated with planets or stars. 
Therefore, all relevant observables are unchanged.
The case of massless test particles, e.g., photons in lensing probes, 
is even simpler; 
the fact that $ds=0$ renders their geodesics invariant 
to Weyl transformations.

In the present framework particle masses scale $\propto\Omega^{-1}$. Since the 
velocity field is subject to the constraint $u_{i}u^{i}=-1$ (summation over spatial indices only) 
then $u^{i}\rightarrow\Omega^{-1}u^{i}$. Consequently, the momentum of massive particles 
$p^{i}=mu^{i}$ scales $\propto\Omega^{-2}$ under Weyl transformation. 
In comparison, the momentum of massless particles 
$p^{i}=u^{i}\rightarrow u^{i}\Omega^{-1}$. These scalings are completely 
analogous to the decay of momenta of massive and massless particles in 
the Friedmann-Robertson-Walker (FRW) spacetime 
$\propto a^{-2}$ \& $\propto a^{-1}$, respectively. 
This analogy is {\it not} coincidental 
as is explained in [14]. As a result, the dimensionless 
ratio of photon energy to Rydberg constant is independent of $\Omega$. 
This implies that while 
particle masses vary in space and time so are photon momenta and energies, and in exactly the 
same fashion. This guarantees that observed atomic and molecular spectra are universal, 
exactly as in GR \& the SM. 
Another way to see this is from the fact that $F^{\mu\nu}\propto\Omega^{-4}$ 
($F_{\mu\nu}\equiv A_{\mu,\nu}-A_{\nu,\mu}$) and 
consequently the energy density of radiation, like that of any other species, 
scales $\propto\Omega^{-4}$, implying that photon energies scale $\propto\Omega^{-1}$, 
exactly as masses do. The observed redshifting universe 
is consequently consistent with monotonically growing particle masses on the 
largest cosmological scales; while the wavelength of radiation is invariant to 
the redshift, photon energies and momenta are sensitive to it. 
This is essentially 
because in the present framework a universal clock is employed in contrast to 
the standard cosmological treatment that assigns cosmic and conformal times 
to massive and massless particles, respectively [14].

Higgs gravity should not be confused with 
fourth-order Weyl-symmetric theory of gravity [15-23], 
which is described by $\mathcal{I}_{W}=-\alpha_{W}
\int C_{\alpha\beta\gamma\delta}C^{\alpha\beta\gamma\delta}\sqrt{-g}d^{4}x$. 
Here, $C_{\mu\nu\rho\sigma}\equiv R_{\mu\nu\rho\sigma}
+(g_{\mu[\sigma}R_{\rho]\nu}+g_{\nu[\rho}R_{\sigma]\mu})+\frac{1}{3}Rg_{\mu[\rho}g_{\sigma]\nu}$, 
where square brackets stand for index anti-symmetrization, 
is defined via the Ricci decomposition of 
the Riemann curvature tensor $R_{\mu\nu\rho\sigma}$ and its contractions, 
and $\alpha_{W}$ is a dimensionless coupling constant. The term 
$C_{\alpha\beta\gamma\delta}C^{\alpha\beta\gamma\delta}$ has conformal weight $-4$, 
and $C^{\alpha}_{\ \beta\gamma\delta}$ has conformal weight $0$, i.e. it is conformally 
invariant.
Possible relation between the two theories at a more fundamental level has been 
discussed in, e.g. [24-27]. Weyl gravity has its own merits [27], a few of them 
might be relevant in the broader context of a combined 
theory of conformal Higgs gravity and Weyl gravity. 
However, for the rest of this work we assume that gravitational dynamics is governed 
only by conformal Higgs gravity as formulated in Eq. (2).

Of fundamental significance in the SM is the SSB mechanism 
by which fundamental particles obtain their masses.
The `wrong' sign of the mass term in Eq. (1) is essential for 
a nonvanishing Higgs VEV in the realm of the SM 
and consequently for the existence of massive particles. 
Since particle masses are universally fixed in the SM 
the Higgs VEV changes discontinuously 
from zero to a finite value at SSB. Within the standard cosmological 
model, this transition is believed to have taken place at 
the Higgs phase transition epoch when the 
typical temperature dropped just below the Higgs scale. 
This epoch is clearly not accessible 
to any direct observation of electromagnetic radiation. Indirect signatures of such 
a phase transition could possibly be imprinted in cosmological data but these, too, 
are expected to be weak and very hard to detect and unequivocally interpret. 
We emphasize that while the 
Higgs particle detection is consistent with the theory described by Eq. (1) -- 
i.e. with its low energy broken-symmetry regime -- it is 
not inconsistent with Eq. (2). The difference between the two is 
due to the fact that there is a phase transition in the SM, whereas there is no such 
transition in the CSM 
because the Higgs VEV 
{\it continuously} 
evolves since particle masses are not bound to be fixed, i.e. energy-momentum is 
not conserved (note Eq. 5). Unless the LHC experiment is repeated at temperatures higher 
than the Higgs scale there is no experimental evidence that 
the SSB mechanism is a process that actually takes place in the SM. 
Thus, our treatment here is limited to the case $R<0$, 
which would correspond to the `broken-symmetry phase', in SM parlance.

\section{Horizons and Singularities}

In general, the fundamental scalar and metric fields of the CSM, Eq. (2), 
are singular. The zeros of these fields play a special role in the proposed 
framework; they define `horizons'. 
More specifically, a horizon associated with the metric field 
represents a limit on its applicability in space or time, 
e.g. Schwarzschild or Kerr horizons. 
This horizon is of the `gravitational-type'. 
Similarly, in the cosmological case the VEV of the Higgs field vanishes at a finite conformal 
time in the past (as does the scale-factor in standard cosmology); this defines the cosmic horizon, 
which is of the `inertial-type' [14]. Unlike a metric-type horizon that 
can be removed by applying an appropriate {\it singular} coordinate transformation, 
an inertial-type horizon cannot be mitigated -- it is a genuine horizon. 

It is widely accepted that gravitation is a fundamental 
interaction, for which reason (and others) it is thought to be an 
inherently quantizable interaction. 
Comparison of Eqs. (1) \& (2) suggests that $R$ is unquantized -- 
it is the effective VEV squared of the 
Higgs field.

Both $R$ and $R_{\alpha\beta\gamma\delta}R^{\alpha\beta\gamma\delta}$ 
(where $R_{\alpha\beta\gamma\delta}$ is the Riemann tensor), 
which are scalars under coordinate transformations, 
are not blind to Weyl transformations 
and can thus be transformed to non-singular functions (or even made 
identically vanishing) everywhere, in analogy to the fact that the 
metric-connection is not a tensor and can thus be made to {\it locally} vanish in GR. 
The difference is that $\Omega(x)$ is a spacetime function and can be so 
chosen, in the CSM framework, 
to regularize the curvature {\it everywhere} in spacetime, not only locally, 
{\it already at the classical level} 
with no recourse to quantum gravity.
Curvature invariants are scalar functions and $\Omega(x)$ 
is just sufficient for this purpose.
This procedure comes at the cost of introducing new singularities in 
the Higgs field, thereby replacing a singularity of the 
gravitational-type with an inertial-type singularity. 

As alluded to above, in this work we entertain the possibility that 
gravitation may well be a genuinely {\it classical collective phenomenon} 
-- and indeed, no {\it empirical} 
result seems to clearly suggest otherwise. Since curvature scalars, 
e.g. $R$ \& $R_{\mu\nu\alpha\beta}R^{\mu\nu\alpha\beta}$, 
do not have well-defined conformal weights we define 
the following curvature invariant
\begin{eqnarray}
\eta\equiv (H^{\dagger}H)^{-2}C_{\mu\nu\rho\sigma}C^{\mu\nu\rho\sigma}, 
\end{eqnarray}
which is a scalar under both coordinate and Weyl transformations 
(it has conformal weight $0$), as a 
diagnostic for singularities. Applying this to the conformally-flat 
FRW metric we readily see that the metric is non-singular. 
In the `comoving frame' 
$ds^{2}=d\eta^{2}-\frac{dr^{2}}{1-Kr^{2}}-r^{2}(d\theta^{2}+\sin^{2}\theta d\varphi^{2})$, 
and consequently $R=-6K$ and $C_{\mu\nu\rho\sigma}=0$; the `initial singularity' 
is absorbed in the inverse Higgs field, i.e. Compton wavelengths of massive particles. 
The modulus of the Higgs field evolves in this picture exactly as would the scale 
factor $a(\eta)$ in the standard cosmological model. In particular, 
it vanishes at the initial time singularity.   

Similarly, the singular (vacuum) exterior Schwarzschild metric 
is characterized by the singular Kretschmann invariant 
$R_{\mu\nu\rho\sigma}R^{\mu\nu\rho\sigma}=12r_{s}^{2}/r^{6}$, where $r_{s}$ 
is the Schwarzschild radius of the gravitating body. Since in vacuum 
$R_{\mu\nu\rho\sigma}=C_{\mu\nu\rho\sigma}$, Eq. (7) implies that a Weyl 
transformation with $\Omega\propto r^{-3/2}$ renders the 
Kretschmann invariant non-singular 
in the new frame by abandoning the linear ruler 
(fixed masses) and replacing it with masses scaling $\propto r^{3/2}$, i.e. 
Compton wavelengths that diverge at the origin as 
$\propto r^{-3/2}$. In other words, the divergent curvature at the origin may be viewed as 
due to either a genuinely singular spacetime (i.e. metric field) 
or rather a consequence of employing a system of units which 
diverges at the origin; in these latter units the radius of curvature is effectively 
infinitely small. These two alternative descriptions are {\it classically} equivalent. 
It should be mentioned that the Schwarzschild spacetime (with fixed Higgs scale) 
is the `near field' limit of the metric obtained for a spherically symmetric static 
vacuum configuration [14] in the framework described by Eq. (2), and applying 
conformal transformations to the metric in this particular case is therefore justified.

A potential source of concern could be that the effective low-energy action 
in Eq. (2) would be extended to include  
conformally-invariant higher-curvature terms when quantum fluctuations of 
the non-metric fields are accounted for, e.g. a term of the form 
$\mathcal{I}_{W}=-\alpha_{W}\int C_{\alpha\beta\gamma\delta}
C^{\alpha\beta\gamma\delta}\sqrt{-g}d^{4}x$, 
which is quadratic in the curvature [11], and possibly higher curvature terms, 
e.g. $\propto\int (H^{\dagger}H)^{2-2n}(C_{\alpha\beta\gamma\delta}C^{\alpha\beta\gamma\delta})^{n}
\sqrt{-g}d^{4}x$ or $\propto\int (H^{\dagger}H)^{2-3n}
(C_{\alpha\beta}^{\ \ \gamma\delta}C_{\gamma\delta}^{\ \ \rho\sigma}
C_{\rho\sigma}^{\ \ \alpha\beta})^{n}\sqrt{-g}d^{4}x$, where $n$ is an 
arbitrary integer. Any singular solution of the new field equations should 
be characterized by invariants of the form described by Eq. (7), or similar 
ones such as $(H^{\dagger}H)^{-3}(C_{\alpha\beta}^{\ \ \gamma\delta}
C_{\gamma\delta}^{\ \ \rho\sigma}C_{\rho\sigma}^{\ \ \alpha\beta})$. 
One can then identify the strongest singularity of 
$|C|_{2}\equiv (C_{\alpha\beta\gamma\delta}C^{\alpha\beta\gamma\delta})^{1/2}$ 
or $|C|_{3}\equiv(C_{\alpha\beta}^{\ \ \gamma\delta}C_{\gamma\delta}^{\ \ \rho\sigma}
C_{\rho\sigma}^{\ \ \alpha\beta})^{1/3}$, and appropriately select $\Omega(x)$ to remove 
the singularity in the curvature invariant in a similar 
procedure to that described 
below Eq. (7) in the context of Schwarzschild metric, making 
use of the fact that $|C|_{i}$ so defined transforms $\propto\Omega^{-2}$.    
Perhaps a more effective -- and fundamentally more appealing -- 
way to ameliorate curvature singularities 
induced by higher-order curvature terms is to add a Weyl-invariant 
Lagrange multiplier of the form 
$\xi \int C_{\mu\nu\rho\sigma}C^{\mu\nu\rho\sigma}\sqrt{-g}d^{4}x$ 
to Eq. (2), where $\xi\gg 1$ is a dimensionless coupling parameter which 
guarantees that $|C|^{2}\equiv C_{\mu\nu\rho\sigma}C^{\mu\nu\rho\sigma}$ 
is minimal. Such a term arises naturally when photon and massless fermion loop 
corrections are accounted for, while requiring that the action is still Weyl 
symmetric. The counter-term in four spacetime dimensions in this case [11] 
is of the form $\mathcal{I}_{coun}=
\xi\int C_{\mu\nu\rho\sigma}C^{\mu\nu\rho\sigma}\sqrt{-g}d^{4}x$ 
and $\xi$ diverges at this spacetime dimension. 

\section{Inertial and Gravitational Phenomena}

In our formulation the SM of particle 
physics and GR are both described by 
the CSM, Eq. (2). Whereas GR corresponds to a fixed Higgs 
field VEV, $H=\sqrt{3}m_{p}\approx 4.22\times 10^{18}$ GeV, 
the generalized SM is characterized by a dynamical 
$H$ which is normalized at the experimentally-inferred 
$\approx 125$ GeV {\it on Earth}. 
Except for this difference, all the other fields transform under the Weyl 
transformation as described below Eq. (2) in inertial 
(IU) and gravitational units (GU). For reasons that will become more clear 
below IU are only used within systems, while GU are applicable at the exterior 
of physical systems (or sub-systems).
Since the action is indifferent to 
whether IU or GU are used, and since in both unit systems the metric transforms similarly, 
it follows that $\mathcal{L}_{IU}=\mathcal{L}_{GU}$ because 
at the boundary of the system (with {\it arbitrary} volume size) the two must 
agree while $H$ is either fixed (GU) or dynamical (IU). 
This implies in particular that all Yukawa coupling parameters $\lambda_{y}$, 
implicitly appearing in Eq. (2) via $\lambda_{y}H\bar{\psi}\psi$, are 
rescaled $\propto\Omega^{-1}$ in going from IU to GU. Similarly, 
$\lambda_{SM}\rightarrow\lambda_{SM}\Omega^{-4}$ in transforming from 
IU to GU. Thus, fixing the Higgs VEV at the Planck scale 
comes at a cost -- the dimensionless coupling parameters $\lambda_{y}$ 
\& $\lambda_{SM}$ are now non-universal. This lies at the heart of the 
DE problem and the Higgs stability problem.
It turns out, as we argue in this and next section, 
that GR and the SM (generalized to account for dynamical Higgs VEV) 
are conformally-related 
in the realm of the theory described by Eq. (2). The conformal factor 
that relates them is typically very large for planets, 
stars, galaxies and obviously the entire universe. 
In the next section we explore the far-reaching ramifications of 
the large size of this factor.

There is a significant difference between the 
ways by which $m_{H}$ and $m_{p}$ are empirically 
determined in the SM and GR, respectively. Whereas the former would naively correspond 
in the CSM (Eq. 2) to the non-perturbative part of $R$ [i.e. $R$ is fixed at 
$O(10^{4})GeV^{2}$ on 
Earth], $m_{p}$ is deduced from the dynamics induced on {\it external} test 
particles by weakly gravitating macroscopic systems, described by the perturbed 
Eq. (2) in the weak (gravitational) field limit $\delta R\sim 2\nabla^{2}\Phi$, which 
is generally spacetime-dependent. In other words, whereas $m_{H}$ is deduced in 
the regime of the unperturbed vacuum-like energy domination, $m_{p}$ is inferred 
from small perturbations of clustering baryonic matter over this dominant background, 
e.g. from the gravitational interaction between planets, lead balls, or any other form of 
NR {\it ordinary matter}. 

We also have compelling (albeit indirect) evidence from BBN that $G$ couples 
relativistic particles to spacetime curvature. Basically, $G$ is viewed here as a 
universal conversion 
factor from inertial density units to curvature scales, 
$l_{curv}^{-2}\sim R\sim G\rho_{M}$. However, no DM or DE particles 
have been observed so far, i.e. no inertial mass is associated with them 
to be converted via $G$. Considering DE and DM as 
gravitating is an experimentally unfounded assumption made 
in standard cosmology.
Although the existence of DM and DE has 
been deduced using (what is commonly considered) gravitational proxies, 
it is a fact that none of them have been used in the inference of $G$ -- 
nor could they -- because, again, their mass (if any) is unknown (and as 
we propose below none of these forms of energy are carried by particles).
Since $G$ is non-dynamical it can be absorbed in the normalization of the 
effective energy densities (as inferred from cosmological observations) associated 
with DM and DE, and although DM and DE affect the dynamics of the non-perturbative 
spacetime curvature, e.g. the evolution of the background universe, they by no means 
induce gravitational (in a sense defined below) effects. 
Their sole effect is inertial.

In our quest for a clear distinction between gravitational and inertial phenomena, 
which is clearly relevant to the Weyl-invariant CSM, it is constructive 
to look at gravity in the weak field limit. 
In the linearized gravity approximation applied to scalar 
perturbations the infinitesimal line element is 
$ds^{2}=(1+2\Phi)d\eta^{2}-(1-2\Phi)g_{ij}dx^{i}dx^{j}$, 
where $g_{ij}$ represents the spatial section of the metric 
field (open, flat or closed geometries), and latin indices run over 
spatial coordinates. 
Similarly, in the case of tensor perturbations 
$ds^{2}=d\eta^{2}-(g_{ij}+h_{ij})dx^{i}dx^{j}$.
In this limit $C_{\mu\nu\rho\sigma}=O(\nabla^{2}\Phi)=O(G\delta\rho_{M})$ 
\& $C_{\mu\nu\rho\sigma}=O(\Box h_{ij})=O(G\pi_{ij})$ 
for scalar and tensor metric perturbations, respectively. 
Thus, $C_{\mu\nu\rho\sigma}\neq 0$ describes 
deviations from Minkowski spacetime, 
and the case $C_{\mu\nu\rho\sigma}=0$ corresponds 
to configurations with vanishing matter 
perturbations $\delta\rho_{M}$ and anisotropic stress 
$\pi_{ij}$ in case of scalar and tensor 
perturbations, respectively. This suggests that only inertial phenomena 
are in play in field configurations characterized by $C_{\mu\nu\rho\sigma}=0$.

There are (indeed over-idealistic) nonlinear `gravitational' 
systems whose evolution does not 
depend on $G$ -- the hallmark of the gravitational interaction -- 
which are characterized by conformally-flat metrics.  
For example, integration of the hydrodynamical equations coupled to the Poisson equation 
in the hydrostatic equilibrium case relates the pressure $P_{M}$ at the core of a star and 
(its over-simplified) 
constant energy density profile $\rho_{M}$ by means of the gravitational 
potential $\Phi$, i.e. $P_{M}=\Phi\rho_{M}/2$, with no recourse to $G$. 
The metric describing this system could be brought to a de-Sitter form in its static 
representation (via coordinate transformation of the time coordinate), 
and is thus characterized by a vanishing Weyl tensor. Any realistic description of 
a gravitating system entails increasing density profile towards its core, with a 
geometry which is generally characterized by a non-vanishing Weyl tensor. 
Another example is the evolution of the cosmological background which is 
governed by the Friedmann equation, and its solution $a(\eta)$ 
(the expansion scale factor) 
does not explicitly depend on $G$; by convention $a(\eta_{0})=1$, where 
$\eta_{0}$ is the 
present time. It follows that the `expansion rate', $H(t)\equiv\dot{a}/a$ (and 
consequently cosmological distances) associated with the smooth conformally-flat 
FRW background spacetime, does not explicitly depend on $G$ either. 

From the above criterion that vanishing Weyl tensor is an indication that only 
inertial effects are in play, it follows that, e.g. {\it cosmological 
redshift stemming from the dynamics of the FRW spacetime is an 
inertial phenomenon}; it is caused by the time-dependence of the Higgs 
VEV in a maximally-symmetric static (Minkowski, if  
spatial curvature vanishes) background [14]. 

In this work we argue that the gravitational properties of any system are encapsulated 
in the observable quantity $\Phi$ (e.g. by means of the gravitational redshift it induces)
and its derivatives, with no explicit recourse to $G$. 
In the weak field limit the strength of gravity at a given physical 
system is characterized by the Newtonian potential $\Phi$ which is a measure 
of the escape velocity, $v_{esc}$, from the system, 
$\Phi=-(v_{esc}/c)^{2}$ at sufficiently small velocities, 
and in any case $|\Phi|\approx 1$ 
corresponds to divergence of the metric field, i.e. the horizon. 
Invariance of $\Phi$ under Weyl transformations follows from its definition 
$g_{00}\equiv 1+2\Phi$ and $g_{ii}\equiv 1-2\Phi$; a weakly gravitating 
system remains so in the presence of arbitrary Weyl transformations. 
However, $\Phi$ is not a scalar under coordinate transformations. Thus, 
we are facing the uncomfortable situation that standard `curvature invariants' 
are not invariant to units transformations and $\Phi$ is not invariant under 
coordinate transformations. A truly invariant quantity is given by Eq. (7).

We propose that both DE \& DM are intimately related to the Higgs field 
(that decouples from electromagnetic radiation classically); 
DE is the self-coupling term $\lambda_{SM}(H^{\dagger}H)^{2}$ 
and DM is identified with the spatial part of the kinetic 
term, i.e. $g^{ij}H_{i}^{\dagger}H_{j}$. 
The latter is not invariant to 
coordinate transformations, and it effectively contributes 
a positive energy density to the energy budget of any system. 
If ignored, an effectively {\it positive} 
energy density is missing from the energy budget. 
We emphasize that in observationally deducing $m_{p}$ 
the scalar field (in Eq. 2) is assumed to be fixed, and consequently the kinetic 
term vanishes, i.e. there is no DM, and consistency with observations requires us {\it to 
invoke its existence}.

In the EH action $R-2\Lambda$ is distinguished 
from $\mathcal{L}_{M}$ by the 
coupling parameter $8\pi G$. 
Accounting for the facts that the kinetic term 
$D_{\mu}H^{\dagger}D^{\mu}H$ in Eq. (2) follows from the Weyl transformation law 
of the curvature scalar $R\rightarrow \Omega^{-2}(R-6\Box\Omega)$ 
(and integration by parts in Eq. 2), it is clear that DM 
is decoupled from $G$ as well -- both DM and DE belong to the left hand side 
of (what would be) the Einstein equations in GR terminology. 
Indeed, the presence of DM could be formally accounted for in GR by 
modifying the gravitational potential, 
but DM genuinely induces no contribution to the Weyl tensor [14]; whereas a combined 
coordinate and conformal transformation could in principle lift these 
specific potentials, they cannot make a non-vanishing Weyl tensor to 
vanish or vice versa.
In standard cosmology DM seems to be required for reconciling the 
{\it kinematics} of star clusters within galaxies, 
strong lensing by galaxy clusters, global flatness of space, and the growth of 
the large scale structure in the linear regime with observations. 
However, all these can be shown to be inertial -- 
rather than gravitational -- effects [14]. 

In the following, we will adopt the agnostic approach that {\it only 
ordinary matter (i.e. matter described by the SM) gravitates}.
Cosmologically, this implies that the dynamical time 
$(G\rho)^{-1/2}$ associated with the gravitational interaction at present is $\sim 5$ 
times larger than is usually thought since the baryonic 
content of the universe is $\sim 4\%$ of 
critical density. 
Consequently, there are $O(10^{91})$ background neutrinos 
in the universe rather than $O(10^{89})$ as in standard cosmology. 
While this latter fact is certainly relevant for 
an order of magnitude estimate of the DE that we provide below, 
this has no other practical impact on the standard cosmological model [14].

\section{Other Ramifications}

Central to the present work is the realization 
that gravity and inertia are related in Eq. (2) via a 
conformal transformation $\Omega=O(N^{1/3})$, 
where $N$ is the number of particles that dominate the 
fermionic energy density in a given physical system or sub-system, i.e. $N=N(x)$.
This particular scaling originates in the fact that the Dirac field 
appearing in the SM action, Eq. (1), is normalized 
to a single fermion [more precisely, $(2\pi)^{3}$ such fermions] 
in a given volume at the ground state while that appearing 
in Eq. (2), a multi-particle state, is normalized in GU to 
$N$ fermionic particles in the system volume 
that dominate its energy density in the ground state.
Indeed, the energy density of NR matter 
that sources gravity in e.g. a planet, is given by $\rho_{NR}=mn$, where $m$ is 
the mean baryonic (essentially, the proton) mass, effectively 
$n=\langle\bar{\psi}\psi\rangle$, and the effective Dirac field 
associated with protons or neutrons is then normalized 
$\propto\sqrt{N/V}$, i.e. $\psi$ is expanded 
in multi-particle states (rather than single-particle 
states as in the SM).
Consequently, the choice $\Omega=O(N^{1/3})$ is ideally suited 
for the purpose of unifying the EH and SM actions 
by Eq. (2); the number density 
of fermionic particles transforms under this specific Weyl transformation 
as $n=N/V\rightarrow n\Omega^{-3}\propto 1/V$. 

In the exterior of physical systems, 
i.e. where $H=\sqrt{3}m_{p}=constant$, GU are employed 
which reflect the irrelevance of the number of degrees of 
freedom of which gravitating systems are composed. GU 
are obtained from the standard IU in the limit of a system made up of a single 
effective `particle'. The gravitational attraction 
between two planets, stars, or lead balls, is 
essentially considered as an interaction between two `point particles' rather than 
a collection of particles -- this is the underlying assumption behind the experimental 
inference of $G$. In contrast, within the physical system, $H=H(N(x))$ varies in 
space and time.

Since $H=H(N)$ then, essentially, $\delta H=0$ in the region exterior to a system 
and consequently there are no inertial but rather only gravitational phenomena, 
for which $H=\sqrt{3}m_{p}$. However, both types of phenomena are manifested 
within the system. Therefore, the boundary condition between the interior and 
exterior of an ideal system with well-defined boundaries is $\delta H=0$. 

Relating $m_{p}^{2}/2$, the prefactor of $R$ 
in the EH action, to the corresponding prefactor 
$H^{\dagger}H/6=v^{2}/12=m_{H}^{2}/(24\lambda_{SM})$ in Eq. (2) 
via a conformal transformation $m_{p}\Omega^{-1}(N)=m_{H}$, we obtain that
\begin{eqnarray}
m_{H}/m_{p}=\Omega^{-1}(N)=2\pi\sqrt{12\lambda_{SM}}N^{-1/3}, 
\end{eqnarray}
where $\Omega$ is related to the different normalizations of the Dirac field 
in IU and GU. The different normalizations of the Dirac fields in these 
two unit systems result in the proportionality factor 
$2\pi N^{-1/3}$ that appears in Eq. (8).  

One consequence of the scaling $m_{H}\propto N^{-1/3}$ 
is that the total mass of a gas made of $N$ massive fermionic particles 
is $M\propto N^{2/3}$ in GU, whereas it is $\propto N$ in IU. 
The overall mass normalization of a macroscopic body is of course 
unchanged in going between the unit frames.

Eq. (8) relates (what appears to be) the 
Planck/EW mass hierarchy problem to the fact that there are 
$O(10^{51})$ protons in Earth. 
We emphasize that this comparison is 
consistent only if other terms in Eq. (2) and the EH action are 
consistent under this transformation. While this is easy to guarantee 
in the case that the term $\lambda_{SM}(H^{\dagger}H)^{2}$ 
dominates over $\rho_{M}$ and the kinetic term $H_{\mu}^{\dagger}H^{\mu}$, 
i.e. in systems dominated by the vacuum-like term, this is not so in general; 
in sufficiently complex systems the term $\lambda_{SM}(H^{\dagger}H)^{2}$ 
decays faster than $\rho_{M}$ and $H_{\mu}^{\dagger}H^{\mu}$, and 
Eq. (8) only represents an upper bound on $m_{H}$ because the kinetic term 
is non-negative.

Since in the present framework 
$\rho_{DE}=\lambda_{SM}(H^{\dagger}H)^{2}=m_{H}^{4}/(16\lambda_{SM})$, 
then assuming three species of neutrinos 
($N_{\nu}=1.47\times 10^{91}$ in the observable universe, 
as in [14]) and using Eq. (8), along with the radial 
profile of the Higgs field in our cosmological model [14], 
results in the upper limit $\rho_{DE}<110\ meV^{4}$, 
an upper limit which is about four times larger than the 
observed value on these scales. The suppression factor $N^{-4/3}$ explains 
the $\sim 122$ orders of magnitude gap between the 
Planck density and the inferred $\rho_{DE}$ on horizon scales.

Zero-point fluctuations do not destabilize this value, e.g. 
by adding an overwhelmingly large Planck energy density. Conventionally, 
it is expected that such a correction is at the level 
$\langle\rho\rangle\approx\int_{0}^{m_{p}}k^{3}dk$ 
where the UV cutoff is the Planck scale. A few comments are in order here. 
First, as we saw above, $m_{p}$ is simply the Higgs mass in the limit of a 
system made of a single particle and is suppressed by $O(122)$ orders of magnitude 
for the entire universe. Second, maintaining Weyl invariance of Eq. (2) 
while introducing a dimensional momentum cutoff is impossible. A few 
possible regularization approaches, discussed in e.g. [28] (and references therein), 
essentially replace the momentum cutoff scale by the scalar field itself or 
a function thereof. Accounting for these we calculate the zero-point 
energy density as
\begin{eqnarray}
\langle\rho\rangle=\int_{0}^{\Lambda_{UV}}\frac{2\pi k^{2}dk}{(2\pi)^{3}}
\sqrt{k^{2}+m^{2}}=f(\alpha)m_{H}^{4},
\end{eqnarray}
where $m_{H}=m_{H}(N)$ is the Higgs mass and we assumed for simplicity 
that $\Lambda_{UV}=\alpha m_{H}$, 
and $f(\alpha)\equiv\frac{1}{32\pi^{2}}
(\alpha(1+\alpha^{2})^{1/2}(1+2\alpha^{2})-\ln(1+\alpha(1+\alpha^{2})^{-1/2}))$ 
is $O(10^{-2})$ for a typical $\alpha=1$. Thus, typically, zero-point energy 
will not significantly shift our estimated value for DE on cosmological scales. 
{\it A similar argument on microscopic scales explains away Higgs mass instability.}
The dark energy `problem' and the Higgs mass `instability' have a common cause; 
in calculating the corresponding zero-point fluctuations we tacitly assume 
a single particle state, $N=1$, whereas in practice both effects are inertial and should 
account for the total number of fermionic particles in the universe, and Earth, respectively.

Since the energy (and lagrangian) densities 
of NR and relativistic species scale differently in GU with $n$, i.e. 
$\propto n$ \& $n^{4/3}$ respectively, 
that is $\propto N$ \& $N^{4/3}$ in a given volume, $\rho_{DE}$ 
is independent on $N$, and energy densities are scaled $\propto\Omega^{-4}$ 
in IU, {\it then even though the energy budget 
of a given system might be dominated by relativistic matter, its (sufficiently 
small) subsystems are dominated by NR matter and on yet smaller scales by DE}. 
In other words, although vacuum-like energy might 
dominate on small scales (as for example is the case in planets and stars), 
its relative contribution drops on larger scales. Yet, even on the largest 
cosmological scales it is still comparable to $\rho_{DM}$ \& $\rho_{b}$.

Comparison of the curvature- and vacuum-like terms 
in $\mathcal{L}_{IU}$ \& $\mathcal{L}_{GU}$ results in 
\begin{eqnarray}
m_{H}^{2}=4\sqrt{3\lambda_{SM}}m_{p}m_{\Lambda}, 
\end{eqnarray}
where $m_{H}$ \& $m_{\Lambda}$ and dynamical 
($m_{\Lambda}\equiv(\Lambda/3)^{1/2}$), and 
the (vacuum-like) self-coupling energy of the Higgs 
field, $\lambda_{SM}(H^{\dagger}H)^{2}$, results from 
$\rho_{DE}=3(m_{p}m_{\Lambda})^{2}$.
Combining Eqs. (8) \& (10) we obtain an expression for 
the de-Sitter radius of a physical system with a fermionic 
energy density dominated by $N$ particles
\begin{eqnarray}
r_{\Lambda}=\frac{l_{p}N^{2/3}}{(2\pi)^{2}\sqrt{3\lambda_{SM}}}, 
\end{eqnarray}
where $\Lambda\equiv 3r_{\Lambda}^{-2}$ and $l_{p}=8.101\times 10^{-33}$ 
is the inverse reduced Planck mass. With $N_{\nu}=1.47\times 10^{91}$ 
in the observable universe [14] we obtain a cosmological de-Sitter 
radius of $\sim 2\times 10^{27}$ cm. In Earth $N_{e}=3.57\times 10^{51}$ 
and its de-Sitter radius is $\sim 7.7$ cm. 

While squeezing a system into its de-Sitter 
horizon $r_{\Lambda}\equiv(\Lambda/3)^{-1/2}$ does not affect 
$\rho_{DE}=m_{H}^{4}/(16\lambda_{SM})$ since it is a function of 
$m_{H}$, which depends on $N$ (as in Eq. 8), it does affect $\rho_{NR}$ 
-- the energy density of NR gas of particles -- 
because the latter is $\propto n_{NR}$, which is inversely proportional to 
the system volume. 
Defining $\rho_{NR,r_{\Lambda}}$ to be the energy density 
obtained by squeezing a system of NR particles into its de-Sitter horizon, 
and again making use of Eqs. (8) \& (10) we obtain the universal relation
\begin{eqnarray}
\rho_{p,r_{\Lambda}}/\rho_{DE}=8\pi\lambda_{p}\lambda_{SM}^{1/2}\approx 0.12, 
\end{eqnarray}
where $\propto n_{NR}$ has been replaced by $\propto n_{p}$ since the gravity of 
a system composed of NR gas is dominated by protons (or neutrons), 
and use has been made in $\lambda_{p}\approx 1/185$, the Yukawa coupling 
associated effectively with the proton -- and appears in its effective Yukawa coupling 
term in Eq. (2) -- is a universal value independent of the complexity of the 
system in IU. 
It can be verified by a direct calculation that $\rho_{p}$ \& $\rho_{DE}$ 
on Earth satisfy this relation. On cosmological scales 
$\rho_{b,r_{\Lambda}}/\rho_{DE}\approx 0.068$ (where $\rho_{b}$ 
is essentially the energy density of 
NR protons) deviates from Eq. (11) by $\approx 40\%$, 
most likely because the derivation of Eq. (10) is based 
on an assumption which does not hold on cosmological scales as $\rho_{DE}$ 
constitutes only $70\%$ of the overall energy budget on these scales.

As to DM, the spatial variation of the Higgs VEV from the surface 
of a typical planet or star (with $N\gg 1$) of radius $L$ to $m_{p}$, at the 
center, implies that $\rho_{CDM}\sim (\frac{\partial H}{\partial r})^{2}\sim m_{p}^{2}L^{-2}$. 
Since the Poisson equation for a typical planet or star integrates 
to $\Phi=4\pi G\int\rho_{b}(\bar{x}')/|\bar{x}
-\bar{x}'|d^{3}\bar{x}'\sim\rho_{b}/\rho_{CDM}$, 
it implies that $\Phi\sim\rho_{b}/\rho_{CDM}$ is independent of $G$. Thus, 
{\it the gravitational potential $\Phi$ gauges the relative fraction of $\rho_{b}$ 
in units of $\rho_{DM}$.} 

In the cosmological context we use the fact that 
by definition the escape velocity, as inferred by a hypothetical 
exterior observer, at the cosmic horizon 
should be of order unity in GU, i.e. $\Phi\sim\rho_{b}/\rho_{DM}\lesssim 1$, 
on these scales. Indeed, on cosmological scales 
$\rho_{DM}=|\frac{\partial H}{\partial r}|^{2}\sim m_{p}^{2}/L_{hor}^{2}\sim m_{p}^{2}\Lambda
\sim\rho_{DE}$ in IU, explaining the apparent puzzling fact that $\rho_{DM}$ is of 
the same order of magnitude of $\rho_{DE}$ at the present, and that $\rho_{b}\lesssim\rho_{DM}$ 
in spite of the very different evolution histories of NR matter and 
vacuum-like energy densities [14]. More generally, at any system of particles 
which fills its de-Sitter horizon $1\sim\Phi\sim\rho_{b}/\rho_{DM}$, and 
from Eq. (12), it then follows that in such a system 
$\rho_{DE}$, $\rho_{DM}$ \& $\rho_{b}$ are all comparable. 
It can be also shown that in an homogeneous and isotropic spacetime 
$\rho_{DM}\leq\rho_{DE}$ [14]. Equality is obtained on horizon scales, but since 
$\rho_{DM}$ is observationally deduced mostly from sub-horizon scales it is 
expected that $\rho_{DM}/\rho_{DE}<1$ is inferred from data, which is indeed the case.

In IU \& GU we obtain on Earth that the curvature scale 
is $O(10^{-16})$ cm \& $O(1)$ cm, respectively, due to the scalings 
$\propto\Omega$ \& $\Omega^{2}$, i.e. Eqs. (8) \& (11). 
Indeed, the curvature scale of Earth would have been its horizon, 
had the electromagnetic interaction been negligible. 
Since DM and DE do not couple to gravity 
and $\rho_{b}$ is smaller than that of $\rho_{DE}$ on Earth 
by $O(10^{25})$, we obtain that $l_{curv}=O(10^{12.5})cm$.
This procedure shows that the Higgs mass can be locally 
fixed at its measured value on the 
Earth surface, and the Planck mass can be fixed globally (in Eq. 2) 
-- all this while having a slightly perturbed 
Minkowski metric, as in standard GR.

From the present work perspective, that gravitation is a collective 
phenomenon rather than fundamental interaction, 
the gravitational potential $\Phi$ is the relevant 
(dimensionless) `coupling parameter' of 
gravity, rather than $\alpha_{g}=GMm$ (where $m$ 
is the mass of a test `particle'). 
Whereas the dimensionless scalar $\Phi$ is invariant to Weyl transformations, 
the (contravariant) vector-potentials of the SM interactions are not; 
$A^{\mu}\propto\Omega^{-2}\propto N^{-2/3}$; consequently, 
$\alpha_{g}/\alpha_{EW}\propto \alpha_{g}/\alpha_{QCD}\propto N^{-2/3}\propto (m_{H}/m_{p})^{2}$, 
where $\alpha_{QCD}$ \& $\alpha_{EW}$ are typically $O(10^{-1})-O(10^{-2})$.
We see again that the EW/Planck scale hierarchy is then 
not a fundamental phenomenon but rather 
a property of complex (macroscopic) systems. The 
EW mass scale is suppressed $\propto N^{-1/3}$ with respect to the Planck scale, 
the latter being merely a dimensional reference scale. 
Consequently {\it no Higgs mass instability arises in the first place}. 

The fact that formally the effective dimensionless gravitational 
coupling between two particles of masses $m_{1}$ \& $m_{2}$, $\alpha_{g}=Gm_{1}m_{2}$, 
and its ratio to $\alpha_{e}$ or $\alpha_{QCD}$ is not a fixed universal constant 
as in the SM but rather depends on $N$, is not 
problematic. It would have been a problem had the gravitational interaction between 
fundamental particles been measured, but no such 
interaction has ever been observed and the only evidence we have for the existence of 
gravity comes from macroscopic systems whose gravity 
is described by $\Phi$, i.e. metric perturbations. 

As to cosmic history, in IU the effective number density of particles 
is fixed (space is essentially static, i.e. it is described in the `comoving frame') 
and $\rho_{vac}/\rho_{M}\propto m_{H}^{3}$. 
Similarly, the energy density of radiation, $\rho_{r}$, 
is fixed in these units; therefore, $\rho_{vac}/\rho_{r}\propto m_{H}^{4}$.
This is compatible with the standard cosmological model if $m_{H}$ is replaced with $a$, 
where $a(t)$ is the scale factor of the expanding ($da/dt>0$) FRW cosmological model.
Thus, a dynamical Higgs VEV can replace $a(t)$, 
i.e. space expansion is replaced by  
contraction of the Higgs Compton wavelength, 
thereby accounting for cosmological redshift [14]. 
Finally, the classical horizon, flatness, and primordial relics 
problems associated with the hot big bang scenario 
are addressed within the CSM framework and are discussed elsewhere [14].

\section{Summary}

In this work we explored the tantalizing idea that local scale invariance 
(Weyl symmetry) -- a powerful symmetry principle that 
significantly constrains the nature of fundamental interactions and greatly 
impacts cosmological models 
-- may be an overarching principle governing physical phenomena and interactions, 
i.e. inertia, gravitation, and the three fundamental interactions.
By and large, any viable formulation of the fundamental physical laws 
should {\it ideally} be independent of 
our system of units (at least classically), 
much like it is independent of coordinate systems (the latter reflects 
the observer state relative to the physical system). 
The former symmetry is not a property of either the SM or GR. 

Weyl symmetry may express a very basic (classical) tenet of the known physical 
laws (at least under certain conditions) -- their {\it self-similarity}, i.e. invariance 
to the complexity of the specific physical (sub-) system, 
which can be gauged by 
the effective number of particles that dominates its fermionic energy budget. 
Surprisingly, and to the best of our knowledge, this latter requirement has never 
before been imposed on physical theories. 

Fixing our yardsticks and gauges (e.g. the Higgs scale) on the surface 
of Earth and employing them universally could possibly be the source of key enigmatic 
properties of the standard cosmological model and the SM, 
e.g. curvature singularities, misinterpretation of the notion of Planck mass, 
seemingly fine-tuned DE and Higgs mass, the need to introduce 
CDM, etc.

The resolution proposed here entails a revision of the foundations of the 
system of units employed in our theories of the fundamental interactions; 
these `problems' are simply symptomatic of the large number of protons 
making up Earth (at which our standard rulers have been set), 
and the very much larger number yet of background neutrinos in the observable universe.
Specifically, since the proposed CSM is conformally-related to gravity via 
a local rescaling of any fundamental field of canonical mass dimension $d$ 
by $\propto N^{-d/3}$, where $N$ is the number of 
particles that dominate the fermionic energy budget of the system [e.g. 
$N=O(10^{51})$ and $N=O(10^{91})$ in Earth and and observable 
universe], 
naive expectations in the SM and standard cosmology for the Higgs mass itself, 
or DE, which are inertial rather than gravitational phenomena, depart from the actual 
values by many orders of magnitude.

In the CSM gravitation naturally emerges as a {\it dynamical} interaction from 
merely imposing Weyl symmetry at the classical action level on the SM. 
Gravitation may well be a {\it genuinely classical collective} phenomenon that provides 
a perturbative description of the exterior of a system, i.e. its propagating degrees 
of freedom (as captured by the nonvanishing components of the Weyl tensor) 
as opposed to inertia. Perhaps more precisely, gravitation is associated with 
density inhomogeneities and stress anisotropy. Since the gravitational constant 
$G\sim l_{p}^{2}$ is 
inferred from the exterior field of Earth (and applied universally) 
it encapsulates the irrelevance of the number of fermionic 
particles composing a system for obtaining its 
gravitational potential. 
From the above it is clear that the dynamics of the largest system, our universe, is 
dominated by its inertia rather than gravity, and in particular implies that the universe 
is homogeneous on the largest scales, and inhomogeneous on smaller scales (the latter are observed from their exteriors). While this does not affect the evolution of {\it observables}, 
e.g. the background universe, linear perturbations, etc., it does provide simple solutions to 
a few long-standing cosmological puzzles, and may well explain away a few other [14]. 

It is normally assumed that on cosmological scales 
gravitation is the only fundamental interaction 
at play as the other three are either 
short-range or essentially cancel out. 
However, and as is highlighted in this work, gravitation 
may not be a {\it fundamental interaction}. It is proposed here that gravitational 
and inertial phenomena are two `irreducible' and `orthogonal' aspects 
of spacetime geometry as is captured by the conformally invariant 
part of the Riemann curvature -- the Weyl tensor. 
According to the proposed framework 
{\it inertial and gravitational phenomena are characterized by 
the conformally-flat and conformally-non-flat sectors of the metric}, respectively. 
This characterization is invariant to Weyl transformations. 
In other words, {\it the gravitational sector represents the propagating degrees 
of freedom associated with the spacetime geometry that describes the physical system}. 

Inertia is independent of $m_{p}$ while gravitation does depend on $m_{p}$. 
It is this anomalously huge mass scale (compares to other natural mass scales) 
that lies at the heart of a 
few longstanding, persistent and puzzling features of the standard cosmological model 
and the SM, e.g. the fact that the observationally-inferred 
$\rho_{DE}$ is suppressed by $\sim 122$ orders of magnitude 
compared to the Planck density, the hierarchy problem and Higgs mass instability, etc. 
A fundamental consequence of the approach advocated here is 
that DE is an inertial phenomenon 
which does not depend on the Planck scale, nor is the latter an appropriate UV 
cutoff for radiative corrections to the Higgs mass.

The unification scheme proposed in this work describes all interactions by 
the same action, Eq. (2), but gravitation is essentially a global/collective 
phenomenon observed at the exterior of systems whereas inertia 
is local, i.e. particle masses are obtained by rescaling $m_{p}$ 
by $N^{-1/3}$. These two aspects cannot be described simultaneously by 
the action, and are obtained by invoking GU and IU, respectively. 
The two systems of units are conformally-related.

The already acknowledged staggering simplicity of the universe may be just the precursor 
for a far simpler universe yet, if the laws of physics are indeed self-similar, 
gravitation is a genuinely classical non-fundamental 
phenomenon complementary to the notion of inertia. 
Additionally, 
DE is accounted for by the Higgs potential energy, CDM 
is replaced by the spatially perturbed Higgs VEV, DE and CDM are 
inertial phenomena experienced by observers within a physical system and not by 
exterior observers, the observed large scale homogeneity of the universe directly 
results from being the largest system (and thus observed from within), 
and the cosmic energy budget is fully described by 
Eq. (2). Other important insight from our formulation is the possibility that 
EW and QCD are not unified at high energies, and consequently the cosmic 
`magnetic monopole problem' does not exist in the first place, 
nor any other primordial phase transition took place. 
Finally, the major motivations for supersymmetry, i.e. 
the Higgs instability problem, CDM, and a better convergence of 
the running fundamental coupling constants at the GUT scale, 
are weakened by our alternative theory. 
Other cosmological implications of the CSM are further explored in [14].

\section*{Acknowledgements}

The author is indebted to Yoel Rephaeli for numerous constructive, critical, and 
thought-provoking discussions which were invaluable for this work.

\end{document}